\documentstyle[aps,manuscript,epsfig]{revtex}
\begin{document}
\title{Enhanced Out-of-plane Emission of K$^+$ Mesons \\
observed in Au+Au Collisions at 1 AGeV
}

\author{
Y.Shin$^c$, 
W.Ahner$^a$,
R.~Barth$^a$,
P.~Beckerle$^c$,
D.~Brill$^c$,
M.~Cie\'slak$^e$,
M.~D\c{e}bowski$^a$,
E.~Grosse$^f$,
P.~Koczo\'n$^a$,
B.~Kohlmeyer$^d$,
M.Mang$^a$,
D.~Mi\'skowiec$^a$,
C.~M\"untz$^b$,
H.~Oeschler$^b$,
F.~P\"uhlhofer$^d$,
E.~Schwab$^a$,
R.~Schicker$^a$,
P.~Senger$^a$,
J.~Speer$^d$,
H.~Str\"obele$^c$,
C.~Sturm$^b$,
K.~V\"olkel$^d$,
A.~Wagner$^b$,
W.~Walu\'s$^e$\\
(KaoS Collaboration)\\ 
$^a$ GSI Darmstadt,
$^b$ TU Darmstadt,
$^c$ Univ. Frankfurt,
$^d$ Univ. Marburg,
$^e$ Univ. Krak\'ow,
$^f$ FZ Rossendorf und TU Dresden
}
\maketitle


\begin{abstract}
The azimuthal angular distribution of K$^+$ mesons 
has been measured  in Au + Au collisions at 1 AGeV.
In peripheral and semi-central collisions,
K$^+$ mesons preferentially are emitted  
perpendicular to the reaction plane.
The strength of the azimuthal anisotropy of K$^+$ emission is 
comparable to the one of pions.
No in-plane flow was found for K$^+$ mesons near projectile and target rapidity.

\end{abstract}

\vspace{.5cm}
\newpage
Relativistic nucleus-nucleus collisions provide the  unique 
possibility to study  both the behaviour of nuclear matter at 
high densities and the properties of hadrons
in the dense medium.  In particular the production and propagation 
of strange mesons  
is considered  to be sensitive to in-medium effects.
Theory predicts kaon-nucleon potentials in nuclear matter which 
are repulsive for K$^+$ but attractive for K$^-$ mesons 
\cite{kaplan,brown1,waas}. 
The large K$^-$ production cross section
found in Ni+Ni collisions  \cite{schroeter,barth}
has been regarded as an evidence for a reduced effective mass    
of the K$^-$ meson in  nuclear matter \cite{li_ko_fang,cassing}.

The K$^+$ mesons are expected to be repelled from the nucleons and hence 
the kaon emission pattern should be anticorrelated with 
the collective motion of the  nucleons. 
Indirect evidence for such a behaviour comes from a 
measurement of the average K$^+$ momentum  
in the event plane in Ni+Ni collisions 
at 1.93 AGeV: the directed in-plane flow  found for  
protons and lambdas near target rapidity is absent for kaons \cite{ritman}.
This finding was attributed to a repulsive K$^+$N potential 
which compensates the effect of the directed nucleon flow \cite{li_ko_li}.

At midrapidity the directed flow of nuclear matter vanishes for
symmetry reasons   and  the 
protons, neutrons and light fragments are found to be 
emitted preferentially perpendicular to the reaction plane 
\cite{gutbrod,leifels,brill2}.  
Such an effect was predicted as a hydrodynamical ''squeeze-out'' 
of nuclear matter from the dense reaction zone 
\cite{stoecker}. However, in contrast to the in-plane flow, only
a minor fraction of the participating nucleons
- mainly those with large transverse momenta - 
take part in this off-plane collective motion \cite{brill2}.  
The azimuthal emission pattern of pions was also found to be
strongly anisotropic at midrapidity. Similar to the nucleons, the pions 
are emitted preferentially perpendicular to the reaction plane
\cite{brill1,venema,brill3}. 
This effect is independent of the pion charge and - as for protons - 
most pronounced for large transverse momenta and semi-central collisions.
The pion nonisotropic emission pattern was attributed to shadowing 
effects caused by rescattering off (or absorption by) 
the spectator fragments \cite{bass}.
The K$^+$ mean free
path for rescattering in nuclear matter is substantially longer than the one 
of pions. 
Therefore, much smaller anisotropies of the K$^+$ azimuthal emission pattern 
were expected.     

In this Letter we present evidence for azimuthal anisotropic 
emission of  K$^+$ mesons  measured in Au+Au collisions at 
a beam kinetic energy of 1 AGeV. 
In-medium effects will be most pronounced
in this very heavy collision system. 
The chosen bombarding energy is well below the K$^+$ production threshold  
(E$_{thres}^{NN}$ = 1.58 GeV for free NN collisions). 
Therefore, the kaons can hardly
be produced in first chance NN collisions but are created inside the
dense reaction zone by multistep processes involving more than two 
nucleons. 

The experiment was performed with the kaon spectrometer (KaoS) at the 
heavy ion synchrotron at GSI \cite{senger}. This magnetic spectrometer 
has a large acceptance 
in solid angle and momentum ($\Omega\approx$30 msr, $p_{max}/p_{min}\approx$2). 
The short distance of 5 - 6.5 m from target 
to focal plane minimizes kaon decays in flight. Particle identification
and trigger are based on a measurement of momentum and time-of-flight. 
Background suppression is performed by track reconstruction based on three
large-area multi-wire chambers. Two scintillator hodoscopes are used for 
event characterization. The centrality of the reaction
is determined by the multiplicity of charged particles measured 
with the large-angle hodoscope. This detector consists of 84 modules 
and covers polar emission angles between 12$^o$ and 48$^o$. The 
orientation of the event plane is reconstructed from the azimuthal emission
angles of the charged projectile spectators. These particles are identified 
(up to Z=8) by their energy loss and time-of-flight measured
with the small-angle hodoscope. This detector 
array, which consists of 380 modules,  is positioned 7~m 
downstream from the target and covers polar 
angles from 0.5$^0$ to 11$^0$.

The measurement was performed with a $^{197}$Au beam accelerated to 
an energy of 1 AGeV  with an intensity of about 5$\times10^7$ ions per spill.
Due to the energy loss in the $^{197}$Au target of 
1.93 g/cm$^2$ thickness the beam energy was reduced on the average by 4\%. 
The K$^+$ mesons are measured 
at polar angles of $\Theta_{lab}$=34$^0$, 44$^0$ and 54$^0$   
over a momentum range of 260 $<$ p$_{lab}<$ 1950 MeV/c.
About 25000 K$^+$ mesons have been registered.

In order to determine the particle azimuthal emission angle,   
the reaction plane has to be reconstructed. 
This is done by the transverse
momentum method \cite{daniel}. The orientation of the reaction plane is 
determined for each event 
by the azimuthal direction of the total transverse momentum    
of all spectator particles
detected in the small-angle hodoscope.   
The orientation of the event plane can be determined with an accuracy 
of 36 degrees (standard deviation) for semi-central collisions and 
55 degrees for peripheral and central collisions \cite{brill2}.

Figure 1 shows the azimuthal distribution of
kaons  detected with the spectrometer  
in peripheral (b$\leq$5 fm), semi-central (b=5-10 fm) 
and central collisions (b$\geq$10 fm). 
The angle $\phi$ is relative to the event plane.
The kaons are measured  at normalized rapidities of 0.2$\leq y/y_{proj}<$0.8 
and within a transverse momentum range of 0.2~GeV/c$\leq p_t<$0.8~GeV/c.   
For peripheral and semi-central events  the K$^+$ 
azimuthal distribution  exhibits  clear maxima  
at $\phi=90^o$ and $\phi=-90^o$ corresponding to an enhanced emission
perpendicular to the reaction plane.
For near central collisions the 
event plane is less  well defined and therefore the effect becomes 
less pronounced.

The azimuthal emission pattern can be parameterized by 
N($\phi$)$\propto$ 1+ $a_1$cos$\phi$ + $a_2$cos2$\phi$.
The parameter $a_1$ quantifies the in-plane emission of the particles
parallel ($a_1>$1) or antiparallel ($a_1<$1) to the impact parameter vector,
whereas  $a_2$ stands for an elliptic emission pattern which may be 
aligned with the event plane ($a_2>$0) or oriented perpendicular to the 
event plane  ($a_2<$0). 
The parameters  were determined by a fit to the data and corrected for the 
uncertainty in the reaction plane reconstruction by 
$a'_{1,2}$ = $a_{1,2}$/$<cos2\Delta\phi>$. The values of 
$<cos2\Delta\phi>$ have been determined by a Monte Carlo simulation
and vary between  0.3 for peripheral and central collisions and
0.5 for semi-central collisions (for details see \cite{brill3}).
The results of the fits including the correction are shown in
Fig. 1 (solid lines) and the parameters  are listed in Table 1. 
The strength of the azimuthal anisotropy is given by the 
ratio R which is the number of K$^+$ mesons emitted perpendicular
to the event plane divided by  
the number of K$^+$ mesons emitted parallel to the event plane
(for $a_2<$0): 
$$R =\frac{N(90^o) + N(-90^o)}{N(0^o) + N(180^o)} = \frac{1-a'_2}{1+a'_2}$$
The values of R for  K$^+$ mesons emitted around midrapidity in 
peripheral, 
semi-central and central collisions are given in Table 1.  
Note that these values are corrected
for the resolution of the reaction-plane determination.

Similar values for the azimuthal asymmetry parameter R have been found for 
pion emission in the same reaction \cite{brill1,brill3}. 
Figure 2 shows the ratio R as a function of the  
transverse momentum p$_t$ both for $\pi^+$ and K$^+$ mesons emitted in 
semi-central Au + Au collisions at 1 AGeV around  midrapidity.
The kaon data are grouped into 3 points because of limited statistics. 
Within the error bars, the kaon azimuthal asymmetry parameter R
does not increase with increasing transverse momentum
in contrast to the one for pions.
This implies that the  K$^+$ transverse momentum distributions
will not vary as a function of the azimuthal angle.    
The centrality dependence of R is different for K$^+$ mesons 
(see Fig. 1)  and pions \cite{brill3}: 
the pion azimuthal asymmetry has a maximum around semi-central
collisions whereas for K$^+$ mesons no reduction of R in 
peripheral collisions is observed. The K$^+$ azimuthal 
asymmetry rather increases weakly  
with increasing impact parameter i.e. with increasing size of the 
spectator remnants.  

It is unlikely that the pion and the kaon 
azimuthal asymmetries are both caused by 
pure rescattering on the spectator fragments 
because of the very different mean free paths of K$^+$ mesons and pions 
in nuclear matter.
The total cross section for 
$\pi^+$p scattering with pion momenta of 0.4 - 0.5 GeV/c is 80 - 40 mb 
\cite{landolt}
corresponding to a mean free path of 
$\lambda_{\pi}$ = 0.8 - 1.6 fm in normal nuclear matter. 
In contrast the K$^+$p total cross section is about 12 mb 
for kaon momenta below 1 GeV/c \cite{landolt} resulting in a mean free path
of $\lambda_{K^+} \approx $ 5 fm. 
Indeed, transport models predict a
K$^+$ very small azimuthal anisotropy for  
semi-central (b = 7 fm) Au+Au collisions at 1 AGeV
when considering only K$^+$ rescattering \cite{li_ko_brown}. The result of 
this RBUU calculation is shown in Fig.3 (dashed line) together with the data 
taken around midrapidity  (0.4$\leq y/y_{proj}<$0.6).
On the other hand, the pronounced K$^+$ out-of-plane emission of the 
experimental data is reproduced by the 
calculations if an additional  repulsive in-medium K$^+$N  potential
is taken into account (solid line in Fig.3). 
Similar conclusions are obtained from a QMD calculation \cite{wang}. 

Another experimental evidence for an in-medium change in the 
kaon-nucleon potential was 
predicted to be the disappearance of K$^+$ directed flow into the
reaction plane \cite{li_ko_li}. Related information 
on the in-plane emission of K$^+$ mesons is obtained by dividing
our K$^+$ sample into intervals of rapidity.  
The  K$^+$ azimuthal distributions for semi-central collisions 
near target rapidity (0.2$\leq y/y_{proj}<$0.4),  
midrapidity (0.4$\leq y/y_{proj}<$0.6) and 
projectile rapidity  (0.6$\leq y/y_{proj}<$0.8) 
are parameterized with the function
N($\phi$)$\propto$ 1+ $a_1$cos$\phi$ + $a_2$cos2$\phi$.
The resulting parameters (corrected for the uncertainty in the 
reaction plane reconstruction)  are given in Table 2.
The coefficient $a'_1$ which measures the strength of the in-plane emission
is subject to a systematical  error of 0.06 which is due to the uncertainty of
the beam position at the small-angle hodoscope. Within the
statistical and systematical errors, the $a'_1$ values in Table 2   
are compatible with zero for all rapidity bins. 
We find no signature for the existence  of an enhanced in-plane
emission of K$^+$ mesons near target or projectile rapidity.
This result is in agreement with the absence  of K$^+$ flow as 
measured in a lighter system at higher bombarding energies \cite{ritman}.   
In the case of pions, a small antiflow has been found in Au+Au collisions
\cite{kintner}. This effect was explained by pion rescattering on the 
spectators in the late stage of the collision
\cite{bass2}.

In summary, we have measured K$^+$ triple differential cross sections 
in Au+Au collisions at 1 AGeV for different polar angles.
The K$^+$ azimuthal angular distribution is found to be anisotropic
in peripheral and semi-central collisions.
The kaons are emitted preferentially
perpendicular to the reaction plane.
The K$^+$ azimuthal anisotropy increases weakly with increasing impact
parameter. In contrast to pions, 
no variation of the azimuthal asymmetry with transverse
momentum of the K$^+$ mesons was found within the large experimental errors. 
No in-plane flow of K$^+$ mesons
was observed close to target and projectile rapidity. 

We thank G.Q. Li for providing us with unpublished results of 
his calculations.
This work was supported by the German Federal Government (BMFT), by the
Polish Committee of Scientific Research (Contract No. 2P03B11109) and 
by the GSI fund for University collaborations.

\begin{table}
Table 1:  Results of the fit 
N($\phi$)$\propto$ 1+ $a'_1$cos$\phi$ + $a'_2$cos2$\phi$
to the K$^+$ azimuthal distributions for normalized rapidities
of 0.2$\leq y/y_{proj}<$0.8. 
The coefficients $a'_1$ and $a'_2$ are corrected for the experimental
resolution of the event  plane determination. The values of $a'_1$ 
are subject to an additional systematical error of 0.06.  

\begin{center}
\begin{tabular}{|c|c|c|c|c|}
centrality & $a'_1$ & $a'_2$ & R & $\chi^2$ \\
\hline
peripheral (b$\geq$10 fm)& -0.063$\pm$0.048& -0.256$\pm$0.051 & 1.68$\pm$0.18 & 0.79\\ 
semi-central (5 fm $<$b$<$10 fm) & 0.064$\pm$0.018& -0.219$\pm$0.021 & 1.56$\pm$0.06 & 2.13\\
central (b$\leq$5 fm) & -0.066$\pm$0.014& -0.044$\pm$0.014 & 1.09$\pm$0.03 & 1.5\\
\end{tabular}
\end{center}
\end{table}

\begin{table}
Table 2:  Results of the fit 
N($\phi$)$\propto$ 1+ $a'_1$cos$\phi$ + $a'_2$cos2$\phi$
to the K$^+$ azimuthal distributions for semi-central collisions
and for three ranges of rapidity.
The coefficients $a'_1$ and $a'_2$are corrected for the experimental
resolution of the event plane determination. The values of $a'_1$ 
are subject to an additional systematical error of 0.06.

\begin{center}
\begin{tabular}{|c|c|c|c|c|}
$y/y_{proj} $& $a'_1$ & $a'_2$ & R & $\chi^2$ \\
\hline
0.2-0.4 & 0.084$\pm$0.027& -0.20$\pm$0.028 & 1.48$\pm$0.08 & 2.06\\ 
0.4-0.6 & 0.043$\pm$0.025& -0.257$\pm$0.029 & 1.68$\pm$0.1 & 2.35\\
0.6-0.8 & 0.038$\pm$0.029& -0.174$\pm$0.030 & 1.42$\pm$0.08 & 0.95\\
\end{tabular}
\end{center}
\end{table}

\newpage 
Fig.1: 
K$^+$ azimuthal angular distribution for peripheral (b$\geq$10 fm),
semi-central (b = 5 - 10 fm) and central (b$\leq$5 fm)
Au+Au collisions at 1 AGeV (from top to bottom). 
The data cover normalized rapidities in the interval 
0.2$\leq y/y_{proj}<$0.8 and transverse momenta in the 
interval 0.2 GeV/c$\leq p_t<$0.8 GeV/c. The lines represent fits to the data
(see text).

\vspace{1.cm}
Fig.2: 
Azimuthal anisotropy parameter R as a function of transverse momenta
for pions and K$^+$ mesons measured in semi-central 
Au+Au collisions at 1 AGeV. The data cover normalized rapidities
in the interval
0.2$\leq y/y_{proj}<$0.8 and transverse momenta in the interval
0.2 GeV/c$\leq p_t<$0.8 GeV/c.  

\vspace{1.cm}
Fig.3: 

K$^+$ azimuthal angular distribution measured  in
semi-central (b = 5 - 10 fm) Au+Au collisions at 1 AGeV around midrapidity
(0.4$\leq y/y_{proj}<$0.6) for K$^+$ transverse momenta  
of  0.2 GeV/c$\leq p_t<$0.8 GeV/c. 
The lines represent results of RBUU calculations for an impact parameter 
of b = 7 fm
\protect\cite{li_ko_brown} without  (dashed line)
and with  an in-medium KN potential (solid line).  
Both calculations take into account kaon-nucleon rescattering.

\newpage       

\begin{figure}[h]
\vspace{0.cm}
\hspace{ 0.cm}\mbox{\epsfig{file=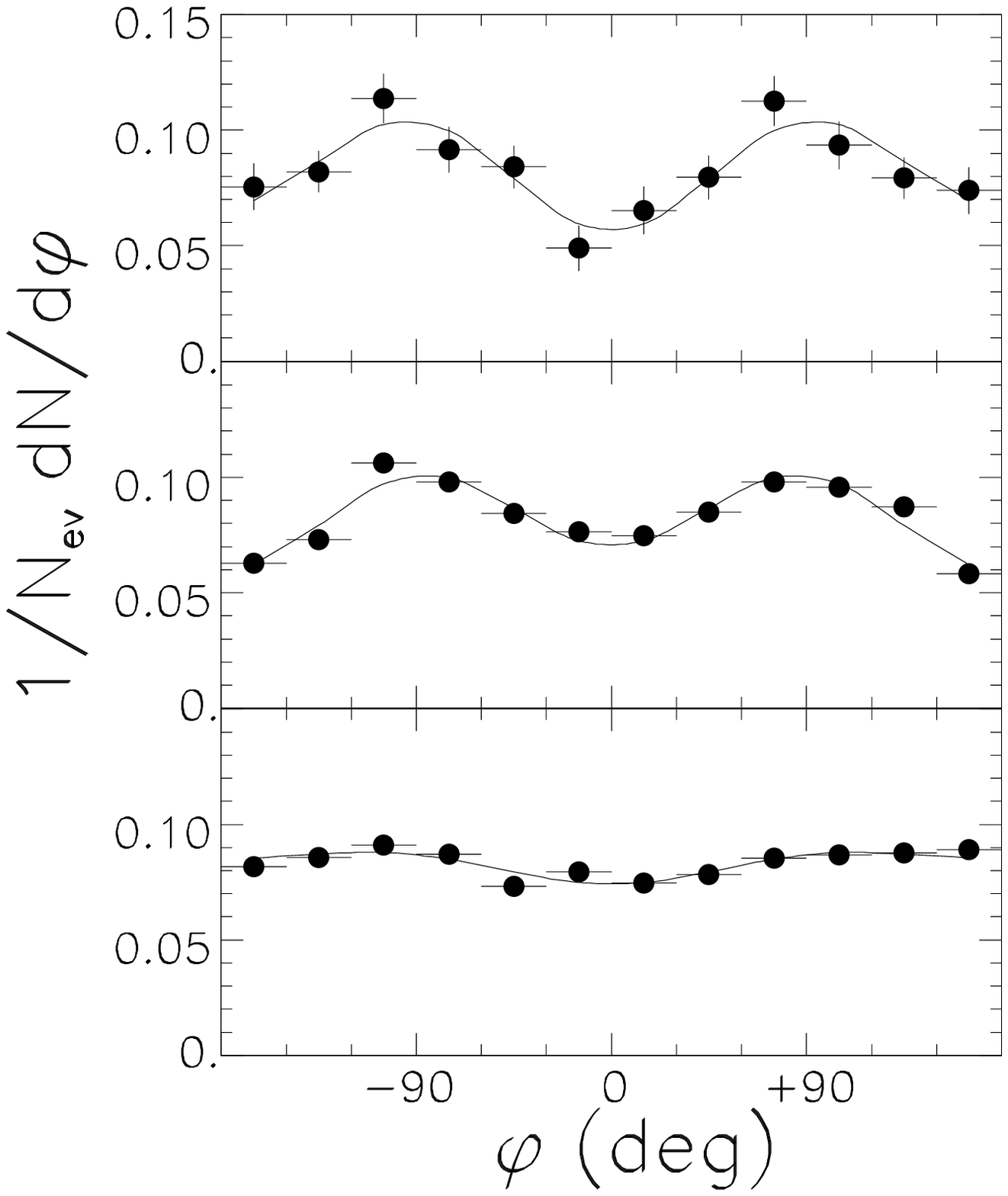,width=14.cm}}
\vspace{.cm}
\caption{}
\end{figure}

\begin{figure}[h]
\vspace{0.cm}
\hspace{ 0.cm}\mbox{\epsfig{file=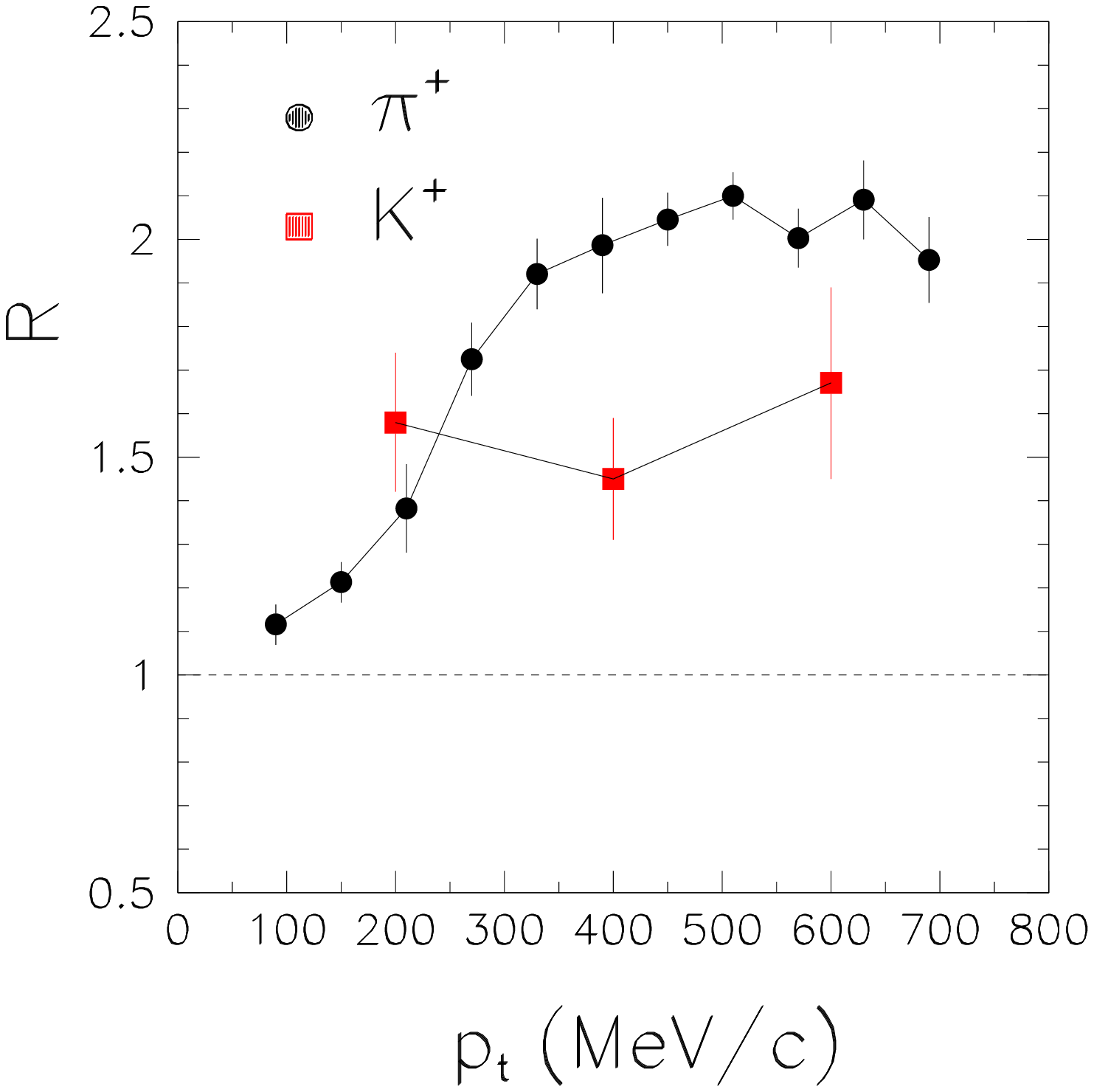,width=16.cm}}
\vspace{.cm}
\caption{}
\end{figure}

\begin{figure}[h]
\vspace{0.cm}
\hspace{ 0.cm}\mbox{\epsfig{file=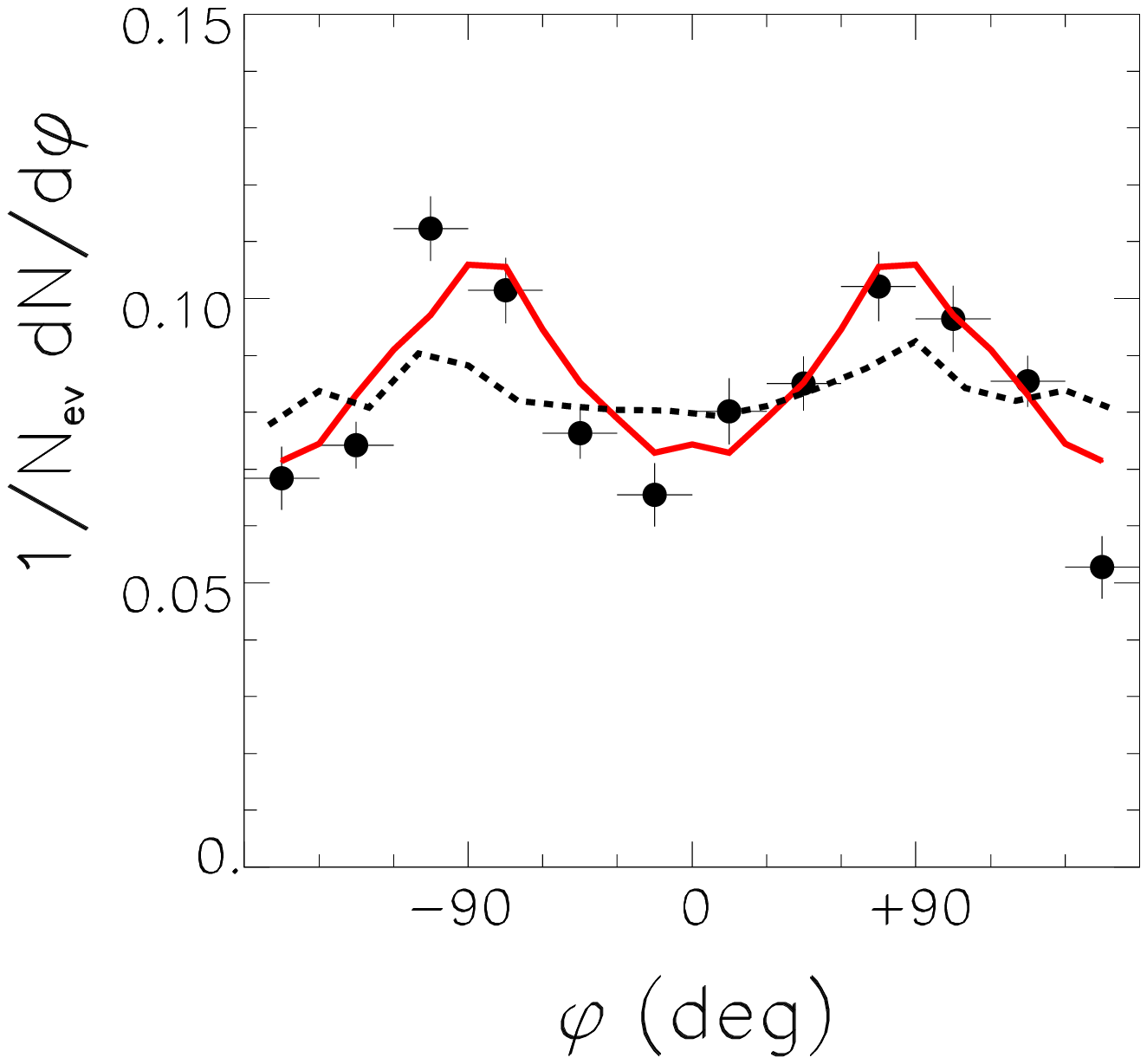,width=18.cm}}
\vspace{.cm}
\caption{}
\end{figure}

\end{document}